\documentclass[aps,prl,twocolumn,superscriptaddress]{revtex4-1}
\usepackage{amsmath}
\usepackage{amssymb}
\usepackage{graphicx}
\usepackage{epstopdf}
\usepackage[colorlinks=true]{hyperref}

\usepackage{physics}
\usepackage{mathrsfs}
\usepackage{comment}

\renewcommand\({\begin{equation}}	
\renewcommand\){\end{equation}}
\begin{document}
\title{Topological Transport of Vorticity in Heisenberg Magnets}

\author{Ji Zou}
\affiliation{Department of Physics and Astronomy, University of California, Los Angeles, California 90095, USA}
\author{Se Kwon Kim}
\affiliation{Department of Physics and Astronomy, University of California, Los Angeles, California 90095, USA}
\affiliation{Department of Physics and Astronomy, University of Missouri, Columbia, Missouri 65211, USA}
\author{Yaroslav Tserkovnyak}
\affiliation{Department of Physics and Astronomy, University of California, Los Angeles, California 90095, USA}

\begin{abstract}
We study a robust topological transport carried by vortices in a thin film of an easy-plane magnetic insulator between two metal contacts. A vortex, which is a nonlocal topological spin texture in two-dimensional magnets, exhibits some beneficial features as compared to skyrmions, which are local topological defects. In particular, the total topological charge carried by vorticity is robust against local fluctuations of the spin order-parameter magnitude. We show that an electric current in one of the magnetized metal contacts can pump vortices into the insulating bulk. Diffusion and two-dimensional nonlocal Coulomb-like interaction between these vortices will establish a steady-state vortex flow. Vortices leaving the bulk produce an electromotive force at another contact, which is related to the current-induced vorticity pumping by the Onsager reciprocity. The voltage signal decays algebraically with the separation between two contacts, similarly to a superfluid spin transport. Finally, the vorticity and closely related skyrmion type topological hydrodynamics are generalized to arbitrary dimensions, in terms of nonsingular order-parameter vector fields.
\end{abstract}

\date{\today}
\maketitle


\textit{Introduction.}|Topology and geometry play an important role in modern condensed matter physics \cite{RevModPhys.51.591, *topology, *RevModPhys.80.1083, *RevModPhys.83.1057}. Topological excitations, which are nonlinear order-parameter textures, are interesting physical objects both theoretically and experimentally \cite{doi:10.1080/00018732.2012.663070, *Parkin190, *M}. Dynamics of these excitations can result in conservation laws that do not result from any symmetries of the system, but rather, derive directly from their topology, rooted in the homotopic properties of the associated fields. A magnetic insulator is a rich platform to study various classes of topological excitations and their (hydro)dynamics. On the practical flip side, we can exploit these excitations to deliver information through charge insulators more effectively than using decaying quasiparticles, such as phonons or magnons \cite{PhysRevLett.109.096603, *nature}. Chiral domain walls in quasi-one-dimensional easy-plane (anti)ferromagnets \cite{doi:10.1080/00018731003739943,PhysRevB.92.220409}, skyrmions in quasi-two-dimensional magnets \cite{PhysRevB.94.024431}, and the winding of three-dimensional spin-glass textures \cite{2018arXiv180301309O} have already been investigated extensively, in this context.

Easy-plane magnets support topological excitations referred to as vortices. They are characterized by the U(1) winding number, similar to superconducting vortices, and thus are nonlocal, being immune to arbitrary local perturbations (or ``surgeries," in the jargon of topologists). This makes them more robust for long-ranged transport than the previously considered topological defects. In addition, their nonlocal nature engenders the Coulomb-like interaction (logarithmic potential), giving rise to a finite-temperature Kosterlitz-Thouless transition. Also, vortices are promising candidates for information and energy storage \cite{2018arXiv180802461T}. In this paper, we will develop the hydrodynamic picture of vortices and realize a superfluid-like transport \cite{doi:10.1080/00018731003739943, PhysRevLett.87.187202, PhysRevLett.112.227201, PhysRevB.90.094408}, based on nonsingular textures in easy-plane magnetic materials.

\begin{figure}
\includegraphics[scale=0.4]{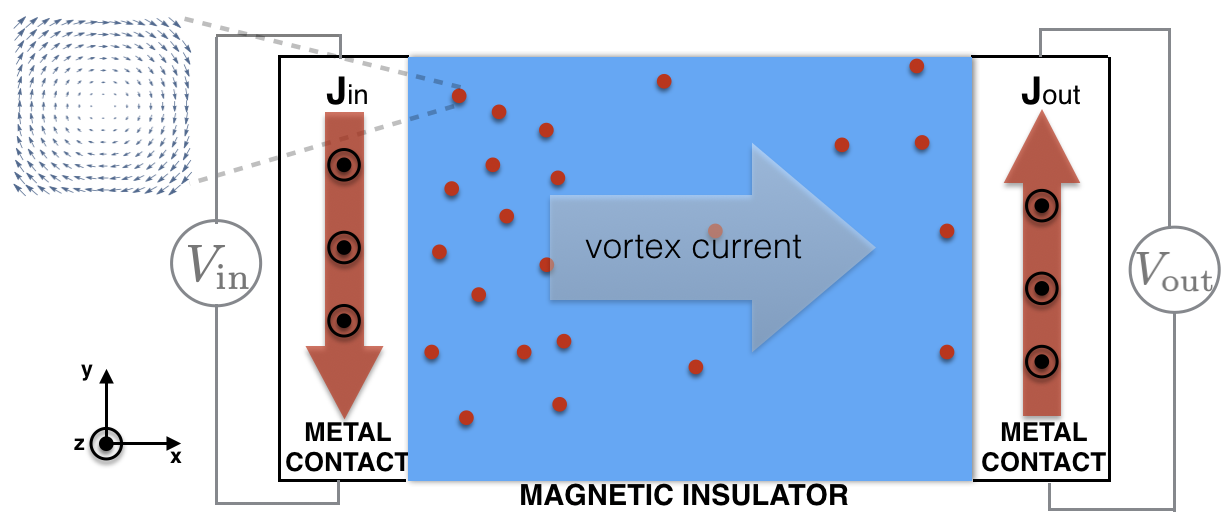}
\caption{A schematic for the proposed injection and detection of vortices. The electric current in the left magnetized contact pumps vortices into the insulating bulk. The applied voltage is $V_\text{in}$.  The vortices leaving the system through the right magnetized contact sustain the output voltage $V_{\text{out}}$. The drag coefficient $\mathcal{C}_d\equiv V_{\text{out}}/V_{\text{in}}$ quantifies the efficiency of the topological vorticity transport.}
\label{fig1}
\end{figure}

\textit{Main results and discussion.}|To illustrate our key findings, we focus on the two-terminal geometry of Fig.~\ref{fig1}. An electric current in the left magnetic metal contact with magnetization $\vb{M}$ exerts an adiabatic torque on the spins of the film at the left boundary. For an appropriate choice of $\vb{M}$ (polarized out of the plane), the work done by the torque will energetically bias the vortex injection into the bulk.   By regarding these vortices as classical objects, diffusion and nonlocal Coulomb interactions \cite{RevModPhys.59.1001} will establish a steady-state distribution of vortex density and its flow. This pumped vorticity will leave the system and induce an electromotive force \cite{0022-3719-20-7-003, *PhysRevB.77.134407, *PhysRevB.80.184411} at the right contact, according to the Onsager-reciprocal process \cite{PhysRev.37.405}. Using the drag coefficient $\mathcal{C}_d\equiv V_{\text{out}}/V_{\text{in}}$ to measure the efficiency of this topological transport, we find
\(\mathcal{C}_d=(\pi \eta M)^2 \sigma_c\sigma\mathcal{A}/L\,,\)
in the linear-response regime, when $L\to\infty$ (so the magnetic-insulator bulk dominates the impedance for the vorticity flow). $\sigma_c$ and $\sigma$ here are the conductivity of electrons in the metal contacts and the effective conductivity of vortices in the insulating bulk, respectively. $\eta$ is a phenomenological parameter measuring the contact efficiency of the charge-vorticity interconversion. $L$ is the length of the magnetic insulator in the $x$ direction, and $\mathcal{A}$ is the cross section of the metal contacts in the $xz$ plane.

A vortex, being a nonlocal spin texture, shows some beneficial features as compared to chiral domain walls and skyrmions. For instance, the total charge of vorticity is robust to local surgery, such as caused by thermal spin fluctuations. We have the same total vorticity charge even if we arbitrarily deform the spin configuration, as long as the changes are local and not emanating to the boundary. In contrast, local fluctuations could be detrimental to domain-wall chirality \cite{PhysRevB.93.020402} and skyrmion number.

At low temperatures, we can generate a vortex lattice in a magnet by utilizing an adiabatic torque on the boundary to control the effective chemical potential associated with the vorticity. This can serve as a platform to explore fundamental physics of emergent solitonic structures, beyond the Abrikosov vortex lattice \cite{ABRIKOSOV1957199} in superconductors or the skyrmion lattice in chiral magnets. Maintaining skyrmionic crystals out of equilibrium, furthermore, can be more challenging, due to their local character (and the associated finite lifetime, when they are metastable).  Another important aspect is the long-ranged Coulombic interactions between the vortices. We may exploit the associated nonlinear effects to realize semiconductor-inspired transport phenomena like \textit{pn} junctions \cite{6773080}. It is also interesting to explore the natural plasma analogies in the ac response.


\textit{Continuity equation and  stability.}|Let us consider a two-dimensional magnetic insulator at low temperatures, such that the coarse-grained local spin-density field $\vb{m}(t,x,y)=(m^x,m^y,m^z)$ captures its low-energy dynamics. The vortex density $\rho$ and flux $\mathbf{j}$ constitute the three-current $j^\mu=(\rho, \vb{j})$ \cite{notation}:
\(j^\mu=\epsilon^{\mu\nu\rho}\epsilon_{zbc} \partial_\nu m^b\partial_\rho m^c/2\pi \,, \label{1}\)
where $a,b,c$ run over three spin-space projections $x,y,z$ and $\mu,\nu,\rho$ run over three time-space coordinates $t,x,y$. It is easy to verify that the density \cite{spinwave} defined in Eq.~(\ref{1}) is conserved: $\partial_\mu j^\mu=0$, so long as the vector field $\vb{m}(t,x,y)$ is smooth such that $\partial_{\mu\nu} \vb{m}=\partial_{\nu\mu} \vb{m} $. This is just the continuity equation: $\partial_t\rho+\div{\vb{j}}=0$.

To see that we can geometrically interpret the current in Eq.~(\ref{1}) as a vortex flow, let us integrate the conserved quantity:
\(\mathcal{Q}=\int_\Omega \rho\,\, dxdy=\frac{1}{2\pi}\int_{\partial\Omega}\vb{m}_{\|}^2  \, \grad{\phi}\cdot  d\vb{l}\,. \)
where we use the Stokes theorem and the fact that $\rho$ is a curl of a vector field. Here, $\vb{m_{\|}}\equiv (m_x,m_y)$ is the planar projection of the vector field, $\phi$ is its polar angle relative to the $x$ axis, and $\Omega$, $\partial\Omega$ denote the bulk and boundary regions. To ensure that $\phi$ is well defined, we should require $m_{\|}\neq 0$. In the case of an easy-plane anisotropy, $m_{\|}=1$ on the boundary away from the vortex core (normalizing the vector field so that $\mathbf{m}\to1$ away from strong textures). Since the polar angle $\phi$ changes by $2\pi \mathcal{Q} $ in one complete anticlockwise passage around the core, we indeed see that Eq.~(\ref{1}) ia a proper expression for the vortex density and current. $\mathcal{Q}$ is simply a $S^1$ winding number.

In the easy-plane limit, the topological robustness is rooted in the map $\vb{m}:S^1\rightarrow S^1$, which can be classified by the fundamental group \cite{homotopy, *geometry} $\pi_1(S^1)=\mathbb{Z}$. The base manifold is $\partial\Omega\simeq S^1$ and $\vb{m}_{\|}$ serves as the $XY$ order parameter (hence $S^1$ target manifold, in the easy-plane case). We can thus see that magnetic textures with different $\mathcal{Q}$ are not smoothly connected to each other, and the total charge, which is completely determined by the boundary configuration, is robust to local surgery. Physically, a vortex is stable because it is a nonlocal object,  which must be moved across the entire system (or towards an antivortex) in order to be eliminated. According to Eq.~(\ref{1}), these topological properties extend to general three-component vector field in two spatial dimensions, even in the absence of quantized vortices.

\begin{figure}
\includegraphics[scale=0.4]{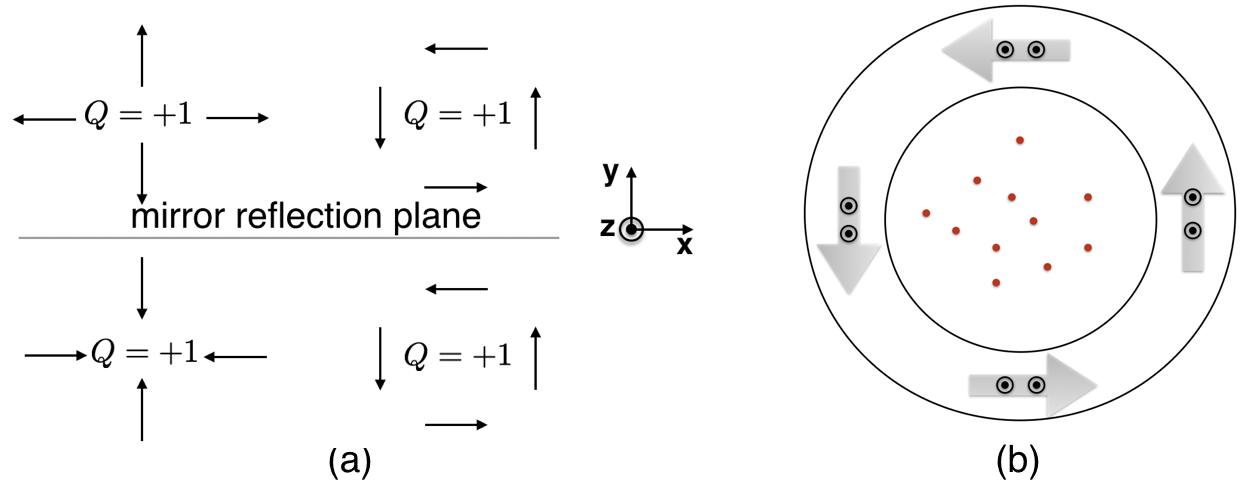}
\caption{Structural symmetries of the vortices and the applied torque setup: (a) Vortex charge $\mathcal{Q}$ is invariant under the $xz$ reflection. (b) A nonequilibrium vortex density can be controlled by the electric current circulating around the magnetic region. Red points represent vortex cores. Arrows represent electric current $\vb{I}=I\hat{\mathbf{t}}$, where the unit vector $\hat{\mathbf{t}}$ curls anticlockwise in the metal contact that is magnetized out of the $xy$ plane.}
\label{fig2}
\end{figure}

\textit{Vortex charge pumping.}|We illustrate the injection of vorticity in Fig.~\ref{fig1}, where the linear electric current density (per unit thickness) $\vb{J}_{\text{in}}=-J_{\text{in}}\hat{\vb{y}}$ in the left contact exerts the local (adiabatic) torque (per unit area in the $yz$ plane) of the form \cite{torque}
\(\vb*{\tau}=\eta\vb{M}\cdot\vb{m}(\vb{J}_{\text{in}}\cdot \nabla)\vb{m} \label{3} \,. \)
$\vb{M}=M\hat{\vb{z}}$ here is the (uniform) out-of-plane magnetization of the metallic contact, $\eta$ is a phenomenological parameter quantifying the strength of the torque, and $\vb{m}$ stands for the (3D) magnetization unit vector  along the interface.  The work done by this torque on the magnetic texture dynamics is then proportional to the vorticity inflow: 
\(W=\int dtdydz \: \vb*{\tau}\cdot (\vb{m}\times \partial_t\vb{m})=\pi \eta M  I_{\text{in}} \mathcal{Q}\,, \)
where $I_{\text{in}}=J_{\text{in}}l$ with  $l$ being the thickness of the system in the $z$ direction and we use the fact $\partial_y\vb{m}\times \partial_t \vb{m}$ is parallel to $\vb{m}$ above. Importantly, the torque discriminates between the topological charges $\mathcal{Q}$ of opposite sign. We denote the work for $\mathcal{Q}=1$ as $W^+\equiv\pi\eta M  I_{\text{in}}$. Note that this work is invariant under the $xz$-plane reflection, which leaves the vortex charge unchanged [see Fig.~\ref{fig2}(a)].

At low temperatures, we can generate a vortex lattice in a magnet by utilizing the torque derived above as follows. In this part, we assume the geometry of the sample is circular [see Fig.~\ref{fig2}(b)].  For  easy-$xy$-plane magnets, now in two spatial dimensions, the energy is given by
\begin{align}
U =&\frac{1}{2}  \int_\Omega dxdy\, \big[ A (\boldsymbol{\nabla} \mathbf{m})^2 + K m_z^2 \big] \nonumber  \\
&-\frac{1}{2} \int_{\partial\Omega} dl \, \eta M \, \hat{\vb{z}} \cdot \big[\mathbf{m} \times (\mathbf{I} \cdot \boldsymbol{\nabla}) \mathbf{m}\big]\,,
\end{align}
where the first term is the bulk energy composed of the exchange energy $\propto A$ and the anisotropy energy $\propto K$, both positive. The second term is the interface energy (integrated over the boundary of the magnet), due to the torque, which is proportional to the net topological charge within the magnet. The current is assumed to flow around the magnetic insulator, tangentially to the boundary: $\mathbf{I} = I \hat{\mathbf{t}}$, with $\hat{\mathbf{t}}$ [see Fig.~\ref{fig2}(b)] defined as the anticlockwise unit vector. If the magnetization lies in the $xy$ plane, in a large $K$ approximation, the energy can be written in terms of the azimuthal angle $\phi$:
\begin{equation}
U = \frac{A}{2} \int_\Omega dxdy \, (\boldsymbol{\nabla} \phi)^2  - \frac{\eta MI }{2}\oint_{\partial\Omega} d\vb{l}\cdot \grad{\phi} \, .
\end{equation}
The second term is quantized as $ -  \pi \eta MI \mathcal{Q}$ where $ \mathcal{Q}$ is the total topological charge. We can minimize the energy
\begin{equation}
U( \mathcal{Q}) = \pi A \mathcal{Q}^2/4 - 2 \pi \eta MI  \mathcal{Q} \label{8}
\end{equation}
with respect to $\mathcal{Q}$, by considering a configuration $\grad{\phi}=\frac{\mathcal{Q}}{R^2}\hat{\vb{z}}\times\vb{r}$ (corresponding to a uniform distribution of vortices), where $R$ is the radius of the sample. The first term $\propto\mathcal{Q}^2$ is the Coulomb interaction energy (which depends on the detailed vortices' distribution) and the second, linear term is the torque-induced energy (which controls the effective ``chemical potential" of the vorticity). The equilibrium winding number for a given current $I$ is thus found to be $ \mathcal{Q}\sim \eta MI /A$. For a fixed $\mathcal{Q}$, the vortices could be expected to form a triangular lattice when $R$ is sufficiently small (depending on the vortex core size $a=\sqrt{A/K}$), in analogy to the Wigner crystal \cite{Wigner:1934aa}. In the opposite regime, as there is no neutralizing background of opposite charge, the vortices should pile up on the edge, which would modify the above electrostatic consideration.


At finite temperatures, similarly to superfluid films, we expect also a Kosterlitz-Thouless transition \cite{KT}, with the critical temperature of $T_{\text{KT}}\sim A/k_B$. When $T>T_{\text{KT}}$, vortex entropy wins over their energetic cost, resulting in the proliferation of vortex pairs. We do not expect the torque-controlled vortex chemical potential to affect the Kosterlitz-Thouless transition in the thermodynamic limit, due to the long-range repulsion of vortices that prevents an extensive build-up of vorticity.

\begin{figure}
\includegraphics[scale=0.3]{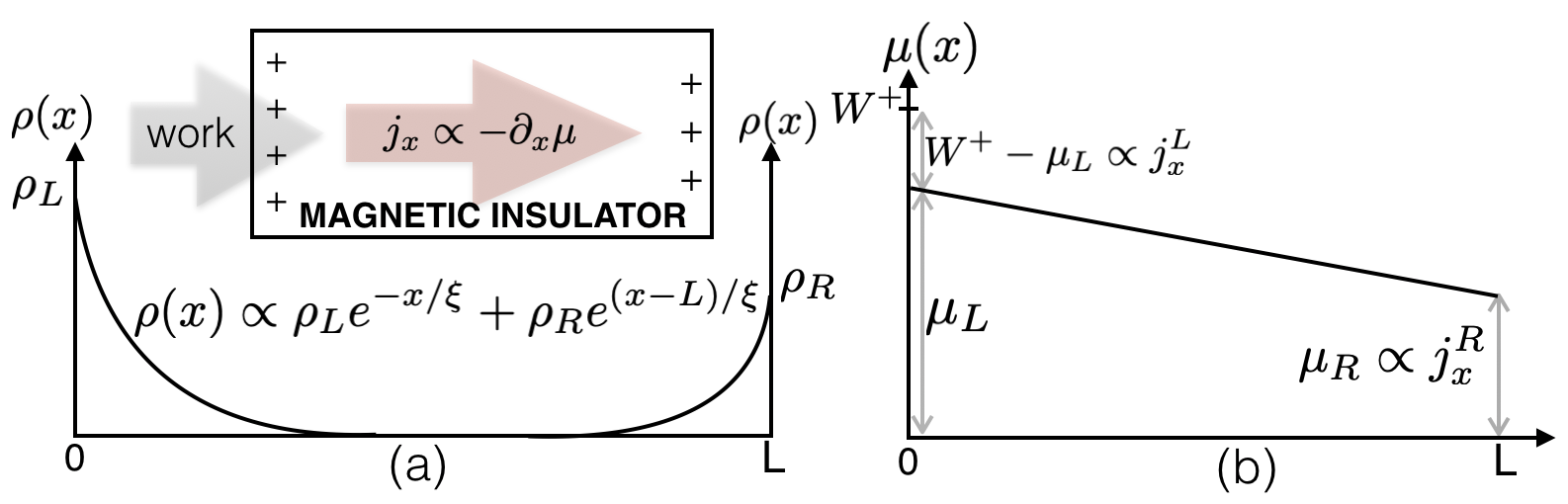}
\caption{(a) Vortex charge density distribution as a function of position $x$, for a two-dimensional strip extended in the transverse direction. The vorticity accumulates near the two ends with the screening length $\xi$. The electric-current induced work $W^+$ (per vortex) at the left end injects vortices into the magnetic bulk. The vortex current  is driven by the electrochemical potential $\mu$. (b) Electrochemical potential $\mu$ as a function of position $x$, in linear response. At the left boundary, the vortex inflow is given by $j_x^L\propto W^+-\mu_L$, while its outflow at the right is $j_x^R\propto \mu_R$. The respective edge electrochemical potentials $\mu_{L,R}$ parametrize the (adiabatic) work required to add an additional vortex there.}
\label{fig3}
\end{figure}

\textit{Topological spin drag.}|In this section, we assume the geometry of the sample is a strip [See Fig.~\ref{fig1}]. Below the temperature $T_{\text{KT}}$, the vortices are bound into neutral pairs, in thermodynamic equilibrium, and the vorticity flow should, therefore, vanish in linear response. Above $T_{\text{KT}}$, the free vortices proliferate, which should result in a finite conductivity $\sigma$. We then expect the constitutive relation $j_x=-\sigma \partial_x \mu$, in terms of the effective electrochemical potential  $\mu=\mu_c+V$. $\mu_c\propto\rho$ here is the chemical potential determined by the local vortex density $\rho$ and $V$ is the electrostatic potential due to the nonlocal Coulomb interaction. The current in the bulk is thus given by 
\begin{align}
j_x=&-\sigma \partial_x \mu= -D\partial_x \rho+\sigma E_x  \,, \label{9}
\end{align}
where $D$ is diffusion coefficient and $\vb{E}=-\grad{V}$ is the fictitious electric field determined by $\div{\vb{E}}=4\pi^2A\rho(x)$, with the (open) exchange boundary conditions (i.e., $E_y=0$ at boundaries). The  coefficient $4\pi^2 A$ is particular for the logarithmic interaction for vortices.  In a steady state, $\partial_xj_x=0$, we obtain charge distribution $\rho(x) \sim  \rho_L e^{-x/\xi}+\rho_R e^{(x-L)/\xi}$ in the bulk, where $\xi\equiv\sqrt{D/4\pi^2 A\sigma}$, when $L\gg\xi$. Vortices accumulate near the two ends on a characteristic lengthscale of $\xi$. The magnetic bulk thus acts like a parallel-plate capacitor [see Fig.~\ref{fig3}(a)]. We can estimate the screening length $\xi$ at high temperatures, $T\gg T_{\text{KT}}$, by treating the vortex plasma as nearly ideal and collisionless. To this end, we invoke the Einstein relation \cite{1905AnP...322..549E}: $D/\sigma=k_BT/\rho_0$, where $\rho_0$ is the equilibrium density of the vortices (irrespective of their charge). This gives $\xi=\sqrt{k_BT/4\pi^2\rho_0A}\sim\sqrt{T/T_{\rm KT}\rho_0}$, which can be interpreted as the Debye-H{\"u}ckel length of our two-dimensional two-component plasma.

From the reaction-rate theory \cite{RevModPhys.62.251}, the vortex inflow at the left boundary is given by
\begin{align}
j_x^L&=\gamma_L^+(T,I) -\gamma_L^-(T,I) \nonumber\\
&=\gamma_L(T)\left[   e^{(W^+-\mu_L)/k_BT}  -   e^{-(W^+-\mu_L)/k_BT}     \right]\nonumber\\
&\approx 2\gamma_L(T)( W^+-\mu_L  )/k_BT\,, \label{11}
\end{align}
in linear response. Here, $\gamma_L^\pm(T,I)$ is the nucleation rate of the vortices with $\mathcal{Q}=\pm1$, in the presence of an applied electric current $I$ [see Fig.~\ref{fig3}(b)]. $\gamma_L(T)\sim\nu_L(T) e^{-E_0/k_BT}$ can be thought of as the equilibrium injection rate of vortices, in terms of the attempt frequency $\nu_L(T)$ and an effective energy barrier $E_0$. Similarly, the vortex outflow at the right boundary is given by \(j^R_x\approx 2\gamma_R(T)\mu_R/k_BT\,, \label{12}\)
which is driven by the electrochemical potential $\mu_R$ that builds up in response to the build up and flow of the vortices from the left contact.

Combining Eqs.~(\ref{9})-(\ref{12}) and imposing $j^L_x=j^R_x=\frac{\sigma}{L}(\mu_L-\mu_R)$,  we find for the steady-state current along the $x$ direction:
\( j_x=\frac{W^+}{ \frac{L}{\sigma}+\frac{k_BT}{2}\left(\frac{1}{\gamma_L}+\frac{1}{\gamma_R}\right) }\,.  \)
The dynamics of the order parameter at the right terminal induces an electromotive force \cite{0022-3719-20-7-003, *PhysRevB.77.134407,  *PhysRevB.80.184411} $ \vb*{\varepsilon}= \pi\eta \vb{j} \times \vb{M} $, according to the Onsager reciprocal relation \cite{PhysRev.37.405}, where $\vb{j}$ is the vorticity outflow normal to the interface and, as before, we are assuming $\vb{M}\propto\hat{\vb{z}}$.
This translates into the induced normalized voltage
\(  \frac{V_{\text{out}}}{V_{\text{in}}} = \frac{ (\pi\eta M)^2  \sigma_c \mathcal{A} }{   \frac{L}{\sigma}+\frac{k_BT}{2}\left(\frac{1}{\gamma_L}+\frac{1}{\gamma_R}\right)  }\,, \label{drag}\)
where $\sigma_c$ is the (Ohmic) conductivity of the metal contacts and $\mathcal{A}$ their cross section in the $xz$ plane. This (negative) drag coefficient between the two metal contacts, which is mediated by the vorticity flow in the magnetic insulator, scales algebraically $\propto L^{-1}$ when $L\to\infty$, which is a generic feature of topological hydrodynamics \cite{PhysRevLett.112.227201,PhysRevB.94.024431}.

In the limit of narrow metal contacts (in the $x$ direction) and a strong magnetic proximity effect due to the magnetic insulator, we can estimate $\eta M\sim\hbar/e$ (in analogy to the adiabatic torques that were invoked in Ref.~\cite{PhysRevB.94.024431} for the generation of skyrmion hydrodynamics). Note, however, that for wider contacts, $\eta$ will scale inversely with their thickness, according to the definition \eqref{3}, so the drag \eqref{drag} will ultimately vanish as $\mathcal{A}^{-1}$. The vorticity hydrodynamics is also expected to get suppressed as the system is scaled up along the $z$ axis. In this limit, as the vortices become larger, $T_{\rm KT}$ increases, the vortex mobility diminishes, while their pinning tendency increases.


\textit{Generalization to higher dimensions.}|Let us consider a $\sigma$ model with symmetry $\text{O}(n+1)$, denoting the order parameter field by $\vb{m}=(m^1,\cdots, m^{n+1})$, in $d$ spatial dimensions. There are two types of topological excitations, in general. The first type is similar to skyrmions, where we collect the infinity at one point, such that the base manifold becomes $S^d$. The order parameter lives in the coset $\text{O}(n+1)/\text{O}(n)\simeq S^n$ since a vectorial order parameter breaks the symmetry $\text{O}(n+1)$ down to $\text{O}(n)$ when the direction of $\vb{m}$ is specified. Topologically-distinct textures are then classified by $\pi_{d}(S^n)$. When $d=n$, we get skyrmionic textures, according to $\pi_{n}(S^n)=\mathbb{Z}$. The associated skyrmion current is given by  \cite{homotopy, *geometry}
\(J^\mu=\epsilon^{\mu\nu_1 \cdots \nu_n}\epsilon_{a_1\cdots a_{n+1}}  m^{a_1} \partial_{\nu_1} m^{a_2} \partial_{\nu_2} m^{a_3} \cdots   \partial_{\nu_n} m^{a_{n+1}}\,, \)
where we use the Greek letters for space-time indices and the Roman letters for field indices.

Leaving the boundaries free, in the $d=n$ case, however, results in another type of excitation. It is analogous to the two-dimensional vorticity and thus more robust than skyrmions due to its nonlocality. To understand this, let us switch to the boundary of a region in $\mathbb{R}^d$ as the base manifold. We can also effectively reduce the dimensionality of the order-parameter space by introducing a hard-axis anisotropy: This gives $\vb{m}_{\|}=(m^1,\cdots, m^d)$. Therefore, the order-parameter manifold is now $S^{d-1}$. The possible topological textures are classified according to $\pi_{d-1}(S^{d-1})=\mathbb{Z}$ on the boundary. Such vorticity-type excitations can always be introduced by adding hard axes, as long as $n+1\geq d$. To make this physically meaningful, we need $m_\|\neq 0$ on the boundary.

Let us now explicitly construct the generalization of vorticity hydrodynamics, for $d=n>1$, as suggested by $\pi_{d-1}(S^{d-1})=\mathbb{Z}$.  The higher-dimensional vorticity current, which obeys the continuity equation $\partial_\mu j^\mu_a~\forall a$, is given by
\(j^\mu_{a}= \epsilon^{\mu\nu_1 \cdots \nu_n}\epsilon_{a a_1\cdots a_{n}}  \partial_{\nu_1} m^{a_1} \partial_{\nu_2} m^{a_2} \cdots   \partial_{\nu_n} m^{a_{n}}\,. \label{vj}\)
To reproduce our preceding discussion of the two-dimensional vorticity flow, we take $d=n=2$ and $a=z$. We can, therefore, regard the vorticity density \eqref{1} as the $z$ component of a vector $\vec{\rho}=(j^0_x,j^0_y,j^0_z)$, which is a conserved \emph{three-dimensional} quantity (corresponding physically to different projections of the order-parameter field). This vectorial vorticity $\vec{\rho}$ is an axial vector, which is invariant under spatial inversion and time reversal. Note the vorticity \eqref{vj} and skyrmion current $J^\mu$ are related via $J^\mu=j^\mu_am_a$, in arbitrary dimensions. 

The $d=n=1$ case is special. The corresponding order-parameter field is two-dimensional: $\vb{m}=(m^x,m^y)$. The vorticity density and current are given by
\(j^\mu_a=\epsilon^{\mu\nu}\epsilon_{ab}\partial_\nu m^b\,,\)
where $\mu,\nu$ can be $t,x$ and $a,b$ can be $x,y$. For example, if we choose $a=x$, then $\rho=\partial_x m^y$ and $j^x=-\partial_t m^y$. Note that the total vorticity charge is bounded:
\(|\mathcal{Q}|=\Big|\int^R_L \partial_xm^ydx\Big|=|m^y(R)-m^y(L)|\leq 2\,,\)
which is due to $\pi_0(S^0)=\mathbb{Z}_2$.


\textit{Summary and discussion.}|It is important to note that we avoid the singular treatment \cite{RevModPhys.59.1001}  of vortex density by allowing the order  parameter to come out of plane in the core. As a result, we have a smooth expression for the vortex density \eqref{1} in terms of the order-parameter field. This density is a conserved quantity obeying a continuity equation, which can thus exhibit a hydrodynamic behavior. We construct the torque (\ref{3}) that is able to inject vortices and show the vortices can mediate algebraically-decaying transconductances in electrical circuits. In contrast to winding or skyrmions, vortices are ``charged" spin textures, which endows them with some beneficial features. For example, vortices are robust against any local perturbation resulted from thermal fluctuation \cite{PhysRevB.93.020402}, as different topological sectors are distinct globally. Furthermore, vortices provide a possibility to realize \textit{pn} junctions or diodes \cite{6773080} for spintronic systems, due to their Coulombic interactions, and may offer opportunities for information and energy storage \cite{2018arXiv180802461T}. Another important point is that we do not require the magnitude of the order parameter to be fixed to have a conserved density. This is again in contrast to winding and skyrmions, where the hydrodynamic picture breaks down when there are strong fluctuations in the order-parameter magnitude.  In this sense, the vortex transport is more stable than other types of spin and topological flows. 

Finally, we remark that our phenomenology of two-dimensional vortex hydrodynamics applies equally well to the antiferromagnetic as well as ferromagnetic films. This is understood from the fact that all the pertinent expressions for the vorticity current, torque, work, etc., are even in the magnetic order parameter $\mathbf{m}$. In a collinear bipartite antiferromagnet, the corresponding contributions from the two sublattices can thus effectively add up, resulting in the same phenomenology.

\begin{acknowledgments}
We are grateful to Hector Ochoa for helpful discussions. The work was supported in part by the NSF under Grant No. DMR-1742928 and the ARO under Contract No. W911NF-14-1-0016.
\end{acknowledgments}


\begin{thebibliography}{34}%
\makeatletter
\providecommand \@ifxundefined [1]{%
 \@ifx{#1\undefined}
}%
\providecommand \@ifnum [1]{%
 \ifnum #1\expandafter \@firstoftwo
 \else \expandafter \@secondoftwo
 \fi
}%
\providecommand \@ifx [1]{%
 \ifx #1\expandafter \@firstoftwo
 \else \expandafter \@secondoftwo
 \fi
}%
\providecommand \natexlab [1]{#1}%
\providecommand \enquote  [1]{``#1''}%
\providecommand \bibnamefont  [1]{#1}%
\providecommand \bibfnamefont [1]{#1}%
\providecommand \citenamefont [1]{#1}%
\providecommand \href@noop [0]{\@secondoftwo}%
\providecommand \href [0]{\begingroup \@sanitize@url \@href}%
\providecommand \@href[1]{\@@startlink{#1}\@@href}%
\providecommand \@@href[1]{\endgroup#1\@@endlink}%
\providecommand \@sanitize@url [0]{\catcode `\\12\catcode `\$12\catcode
  `\&12\catcode `\#12\catcode `\^12\catcode `\_12\catcode `\%12\relax}%
\providecommand \@@startlink[1]{}%
\providecommand \@@endlink[0]{}%
\providecommand \url  [0]{\begingroup\@sanitize@url \@url }%
\providecommand \@url [1]{\endgroup\@href {#1}{\urlprefix }}%
\providecommand \urlprefix  [0]{URL }%
\providecommand \Eprint [0]{\href }%
\providecommand \doibase [0]{http://dx.doi.org/}%
\providecommand \selectlanguage [0]{\@gobble}%
\providecommand \bibinfo  [0]{\@secondoftwo}%
\providecommand \bibfield  [0]{\@secondoftwo}%
\providecommand \translation [1]{[#1]}%
\providecommand \BibitemOpen [0]{}%
\providecommand \bibitemStop [0]{}%
\providecommand \bibitemNoStop [0]{.\EOS\space}%
\providecommand \EOS [0]{\spacefactor3000\relax}%
\providecommand \BibitemShut  [1]{\csname bibitem#1\endcsname}%
\let\auto@bib@innerbib\@empty
\bibitem [{\citenamefont {Mermin}(1979)}]{RevModPhys.51.591}%
  \BibitemOpen
  \bibfield  {author} {\bibinfo {author} {\bibfnamefont {N.~D.}\ \bibnamefont
  {Mermin}},\ }\href {\doibase 10.1103/RevModPhys.51.591} {\bibfield  {journal}
  {\bibinfo  {journal} {Rev. Mod. Phys.}\ }\textbf {\bibinfo {volume} {51}},\
  \bibinfo {pages} {591} (\bibinfo {year} {1979})}\BibitemShut {NoStop}%
\bibitem [{\citenamefont {Thouless}(1998)}]{topology}%
  \BibitemOpen
  \bibfield  {author} {\bibinfo {author} {\bibfnamefont {D.}~\bibnamefont
  {Thouless}},\ }\href@noop {} {\emph {\bibinfo {title} {Topological Quantum
  Numbers in Nonrelativistic Physics}}}\ (\bibinfo  {publisher} {World
  Scientific Publishing Company},\ \bibinfo {year} {1998})\BibitemShut
  {NoStop}%
\bibitem [{\citenamefont {Nayak}\ \emph {et~al.}(2008)\citenamefont {Nayak},
  \citenamefont {Simon}, \citenamefont {Stern}, \citenamefont {Freedman},\ and\
  \citenamefont {Das~Sarma}}]{RevModPhys.80.1083}%
  \BibitemOpen
  \bibfield  {author} {\bibinfo {author} {\bibfnamefont {C.}~\bibnamefont
  {Nayak}}, \bibinfo {author} {\bibfnamefont {S.~H.}\ \bibnamefont {Simon}},
  \bibinfo {author} {\bibfnamefont {A.}~\bibnamefont {Stern}}, \bibinfo
  {author} {\bibfnamefont {M.}~\bibnamefont {Freedman}}, \ and\ \bibinfo
  {author} {\bibfnamefont {S.}~\bibnamefont {Das~Sarma}},\ }\href {\doibase
  10.1103/RevModPhys.80.1083} {\bibfield  {journal} {\bibinfo  {journal} {Rev.
  Mod. Phys.}\ }\textbf {\bibinfo {volume} {80}},\ \bibinfo {pages} {1083}
  (\bibinfo {year} {2008})}\BibitemShut {NoStop}%
\bibitem [{\citenamefont {Qi}\ and\ \citenamefont
  {Zhang}(2011)}]{RevModPhys.83.1057}%
  \BibitemOpen
  \bibfield  {author} {\bibinfo {author} {\bibfnamefont {X.-L.}\ \bibnamefont
  {Qi}}\ and\ \bibinfo {author} {\bibfnamefont {S.-C.}\ \bibnamefont {Zhang}},\
  }\href {\doibase 10.1103/RevModPhys.83.1057} {\bibfield  {journal} {\bibinfo
  {journal} {Rev. Mod. Phys.}\ }\textbf {\bibinfo {volume} {83}},\ \bibinfo
  {pages} {1057} (\bibinfo {year} {2011})}\BibitemShut {NoStop}%
\bibitem [{\citenamefont {Braun}(2012)}]{doi:10.1080/00018732.2012.663070}%
  \BibitemOpen
  \bibfield  {author} {\bibinfo {author} {\bibfnamefont {H.-B.}\ \bibnamefont
  {Braun}},\ }\href {\doibase 10.1080/00018732.2012.663070} {\bibfield
  {journal} {\bibinfo  {journal} {Adv. Phys.}\ }\textbf {\bibinfo {volume}
  {61}},\ \bibinfo {pages} {1} (\bibinfo {year} {2012})}\BibitemShut {NoStop}%
\bibitem [{\citenamefont {Parkin}\ \emph {et~al.}(2008)\citenamefont {Parkin},
  \citenamefont {Hayashi},\ and\ \citenamefont {Thomas}}]{Parkin190}%
  \BibitemOpen
  \bibfield  {author} {\bibinfo {author} {\bibfnamefont {S.~S.~P.}\
  \bibnamefont {Parkin}}, \bibinfo {author} {\bibfnamefont {M.}~\bibnamefont
  {Hayashi}}, \ and\ \bibinfo {author} {\bibfnamefont {L.}~\bibnamefont
  {Thomas}},\ }\href
  {http://science.sciencemag.org/content/320/5873/190.abstract} {\bibfield
  {journal} {\bibinfo  {journal} {Science}\ }\textbf {\bibinfo {volume}
  {320}},\ \bibinfo {pages} {190} (\bibinfo {year} {2008})}\BibitemShut
  {NoStop}%
\bibitem [{\citenamefont {M{\"u}hlbauer}\ \emph {et~al.}(2009)\citenamefont
  {M{\"u}hlbauer}, \citenamefont {Binz}, \citenamefont {Jonietz}, \citenamefont
  {Pfleiderer}, \citenamefont {Rosch}, \citenamefont {Neubauer}, \citenamefont
  {Georgii},\ and\ \citenamefont {B{\"o}ni}}]{M}%
  \BibitemOpen
  \bibfield  {author} {\bibinfo {author} {\bibfnamefont {S.}~\bibnamefont
  {M{\"u}hlbauer}}, \bibinfo {author} {\bibfnamefont {B.}~\bibnamefont {Binz}},
  \bibinfo {author} {\bibfnamefont {F.}~\bibnamefont {Jonietz}}, \bibinfo
  {author} {\bibfnamefont {C.}~\bibnamefont {Pfleiderer}}, \bibinfo {author}
  {\bibfnamefont {A.}~\bibnamefont {Rosch}}, \bibinfo {author} {\bibfnamefont
  {A.}~\bibnamefont {Neubauer}}, \bibinfo {author} {\bibfnamefont
  {R.}~\bibnamefont {Georgii}}, \ and\ \bibinfo {author} {\bibfnamefont
  {P.}~\bibnamefont {B{\"o}ni}},\ }\href
  {http://science.sciencemag.org/content/323/5916/915.abstract} {\bibfield
  {journal} {\bibinfo  {journal} {Science}\ }\textbf {\bibinfo {volume}
  {323}},\ \bibinfo {pages} {915} (\bibinfo {year} {2009})}\BibitemShut
  {NoStop}%
\bibitem [{\citenamefont {Zhang}\ and\ \citenamefont
  {Zhang}(2012)}]{PhysRevLett.109.096603}%
  \BibitemOpen
  \bibfield  {author} {\bibinfo {author} {\bibfnamefont {S.~S.~L.}\
  \bibnamefont {Zhang}}\ and\ \bibinfo {author} {\bibfnamefont
  {S.}~\bibnamefont {Zhang}},\ }\href {\doibase 10.1103/PhysRevLett.109.096603}
  {\bibfield  {journal} {\bibinfo  {journal} {Phys. Rev. Lett.}\ }\textbf
  {\bibinfo {volume} {109}},\ \bibinfo {pages} {096603} (\bibinfo {year}
  {2012})}\BibitemShut {NoStop}%
\bibitem [{\citenamefont {Kajiwara}\ \emph {et~al.}(2010)\citenamefont
  {Kajiwara}, \citenamefont {Harii}, \citenamefont {Takahashi}, \citenamefont
  {Ohe}, \citenamefont {Uchida}, \citenamefont {Mizuguchi}, \citenamefont
  {Umezawa}, \citenamefont {Kawai}, \citenamefont {Ando}, \citenamefont
  {Takanashi}, \citenamefont {Maekawa},\ and\ \citenamefont {Saitoh}}]{nature}%
  \BibitemOpen
  \bibfield  {author} {\bibinfo {author} {\bibfnamefont {Y.}~\bibnamefont
  {Kajiwara}}, \bibinfo {author} {\bibfnamefont {K.}~\bibnamefont {Harii}},
  \bibinfo {author} {\bibfnamefont {S.}~\bibnamefont {Takahashi}}, \bibinfo
  {author} {\bibfnamefont {J.}~\bibnamefont {Ohe}}, \bibinfo {author}
  {\bibfnamefont {K.}~\bibnamefont {Uchida}}, \bibinfo {author} {\bibfnamefont
  {M.}~\bibnamefont {Mizuguchi}}, \bibinfo {author} {\bibfnamefont
  {H.}~\bibnamefont {Umezawa}}, \bibinfo {author} {\bibfnamefont
  {H.}~\bibnamefont {Kawai}}, \bibinfo {author} {\bibfnamefont
  {K.}~\bibnamefont {Ando}}, \bibinfo {author} {\bibfnamefont {K.}~\bibnamefont
  {Takanashi}}, \bibinfo {author} {\bibfnamefont {S.}~\bibnamefont {Maekawa}},
  \ and\ \bibinfo {author} {\bibfnamefont {E.}~\bibnamefont {Saitoh}},\ }\href
  {http://dx.doi.org/10.1038/nature08876} {\bibfield  {journal} {\bibinfo
  {journal} {Nature}\ }\textbf {\bibinfo {volume} {464}},\ \bibinfo {pages}
  {262} (\bibinfo {year} {2010})}\BibitemShut {NoStop}%
\bibitem [{\citenamefont {Sonin}(2010)}]{doi:10.1080/00018731003739943}%
  \BibitemOpen
  \bibfield  {author} {\bibinfo {author} {\bibfnamefont {E.~B.}\ \bibnamefont
  {Sonin}},\ }\href {\doibase 10.1080/00018731003739943} {\bibfield  {journal}
  {\bibinfo  {journal} {Adv. Phys.}\ }\textbf {\bibinfo {volume} {59}},\
  \bibinfo {pages} {181} (\bibinfo {year} {2010})}\BibitemShut {NoStop}%
\bibitem [{\citenamefont {Kim}\ \emph {et~al.}(2015)\citenamefont {Kim},
  \citenamefont {Takei},\ and\ \citenamefont
  {Tserkovnyak}}]{PhysRevB.92.220409}%
  \BibitemOpen
  \bibfield  {author} {\bibinfo {author} {\bibfnamefont {S.~K.}\ \bibnamefont
  {Kim}}, \bibinfo {author} {\bibfnamefont {S.}~\bibnamefont {Takei}}, \ and\
  \bibinfo {author} {\bibfnamefont {Y.}~\bibnamefont {Tserkovnyak}},\ }\href
  {\doibase 10.1103/PhysRevB.92.220409} {\bibfield  {journal} {\bibinfo
  {journal} {Phys. Rev. B}\ }\textbf {\bibinfo {volume} {92}},\ \bibinfo
  {pages} {220409} (\bibinfo {year} {2015})}\BibitemShut {NoStop}%
\bibitem [{\citenamefont {Ochoa}\ \emph {et~al.}(2016)\citenamefont {Ochoa},
  \citenamefont {Kim},\ and\ \citenamefont {Tserkovnyak}}]{PhysRevB.94.024431}%
  \BibitemOpen
  \bibfield  {author} {\bibinfo {author} {\bibfnamefont {H.}~\bibnamefont
  {Ochoa}}, \bibinfo {author} {\bibfnamefont {S.~K.}\ \bibnamefont {Kim}}, \
  and\ \bibinfo {author} {\bibfnamefont {Y.}~\bibnamefont {Tserkovnyak}},\
  }\href {\doibase 10.1103/PhysRevB.94.024431} {\bibfield  {journal} {\bibinfo
  {journal} {Phys. Rev. B}\ }\textbf {\bibinfo {volume} {94}},\ \bibinfo
  {pages} {024431} (\bibinfo {year} {2016})}\BibitemShut {NoStop}%
\bibitem [{\citenamefont {Ochoa}\ \emph {et~al.}(2018)\citenamefont {Ochoa},
  \citenamefont {Zarzuela},\ and\ \citenamefont
  {Tserkovnyak}}]{2018arXiv180301309O}%
  \BibitemOpen
  \bibfield  {author} {\bibinfo {author} {\bibfnamefont {H.}~\bibnamefont
  {Ochoa}}, \bibinfo {author} {\bibfnamefont {R.}~\bibnamefont {Zarzuela}}, \
  and\ \bibinfo {author} {\bibfnamefont {Y.}~\bibnamefont {Tserkovnyak}},\
  }\href {\doibase 10.1103/PhysRevB.98.054424} {\bibfield  {journal} {\bibinfo
  {journal} {Phys. Rev. B}\ }\textbf {\bibinfo {volume} {98}},\ \bibinfo
  {pages} {054424} (\bibinfo {year} {2018})}\BibitemShut {NoStop}%
\bibitem [{\citenamefont {Tserkovnyak}\ and\ \citenamefont
  {Xiao}(2018)}]{2018arXiv180802461T}%
  \BibitemOpen
  \bibfield  {author} {\bibinfo {author} {\bibfnamefont {Y.}~\bibnamefont
  {Tserkovnyak}}\ and\ \bibinfo {author} {\bibfnamefont {J.}~\bibnamefont
  {Xiao}},\ }\href {\doibase 10.1103/PhysRevLett.121.127701} {\bibfield
  {journal} {\bibinfo  {journal} {Phys. Rev. Lett.}\ }\textbf {\bibinfo
  {volume} {121}},\ \bibinfo {pages} {127701} (\bibinfo {year}
  {2018})}\BibitemShut {NoStop}%
\bibitem [{\citenamefont {K{\"o}nig}\ \emph {et~al.}(2001)\citenamefont
  {K{\"o}nig}, \citenamefont {B{\o}nsager},\ and\ \citenamefont
  {MacDonald}}]{PhysRevLett.87.187202}%
  \BibitemOpen
  \bibfield  {author} {\bibinfo {author} {\bibfnamefont {J.}~\bibnamefont
  {K{\"o}nig}}, \bibinfo {author} {\bibfnamefont {M.~C.}\ \bibnamefont
  {B{\o}nsager}}, \ and\ \bibinfo {author} {\bibfnamefont {A.~H.}\ \bibnamefont
  {MacDonald}},\ }\href {\doibase 10.1103/PhysRevLett.87.187202} {\bibfield
  {journal} {\bibinfo  {journal} {Phys. Rev. Lett.}\ }\textbf {\bibinfo
  {volume} {87}},\ \bibinfo {pages} {187202} (\bibinfo {year}
  {2001})}\BibitemShut {NoStop}%
\bibitem [{\citenamefont {Takei}\ and\ \citenamefont
  {Tserkovnyak}(2014)}]{PhysRevLett.112.227201}%
  \BibitemOpen
  \bibfield  {author} {\bibinfo {author} {\bibfnamefont {S.}~\bibnamefont
  {Takei}}\ and\ \bibinfo {author} {\bibfnamefont {Y.}~\bibnamefont
  {Tserkovnyak}},\ }\href {\doibase 10.1103/PhysRevLett.112.227201} {\bibfield
  {journal} {\bibinfo  {journal} {Phys. Rev. Lett.}\ }\textbf {\bibinfo
  {volume} {112}},\ \bibinfo {pages} {227201} (\bibinfo {year}
  {2014})}\BibitemShut {NoStop}%
\bibitem [{\citenamefont {Takei}\ \emph {et~al.}(2014)\citenamefont {Takei},
  \citenamefont {Halperin}, \citenamefont {Yacoby},\ and\ \citenamefont
  {Tserkovnyak}}]{PhysRevB.90.094408}%
  \BibitemOpen
  \bibfield  {author} {\bibinfo {author} {\bibfnamefont {S.}~\bibnamefont
  {Takei}}, \bibinfo {author} {\bibfnamefont {B.~I.}\ \bibnamefont {Halperin}},
  \bibinfo {author} {\bibfnamefont {A.}~\bibnamefont {Yacoby}}, \ and\ \bibinfo
  {author} {\bibfnamefont {Y.}~\bibnamefont {Tserkovnyak}},\ }\href {\doibase
  10.1103/PhysRevB.90.094408} {\bibfield  {journal} {\bibinfo  {journal} {Phys.
  Rev. B}\ }\textbf {\bibinfo {volume} {90}},\ \bibinfo {pages} {094408}
  (\bibinfo {year} {2014})}\BibitemShut {NoStop}%
\bibitem [{\citenamefont {Minnhagen}(1987)}]{RevModPhys.59.1001}%
  \BibitemOpen
  \bibfield  {author} {\bibinfo {author} {\bibfnamefont {P.}~\bibnamefont
  {Minnhagen}},\ }\href {\doibase 10.1103/RevModPhys.59.1001} {\bibfield
  {journal} {\bibinfo  {journal} {Rev. Mod. Phys.}\ }\textbf {\bibinfo {volume}
  {59}},\ \bibinfo {pages} {1001} (\bibinfo {year} {1987})}\BibitemShut
  {NoStop}%
\bibitem [{\citenamefont {Volovik}(1987)}]{0022-3719-20-7-003}%
  \BibitemOpen
  \bibfield  {author} {\bibinfo {author} {\bibfnamefont {G.~E.}\ \bibnamefont
  {Volovik}},\ }\href {http://stacks.iop.org/0022-3719/20/i=7/a=003} {\bibfield
   {journal} {\bibinfo  {journal} {J. Phys. C}\ }\textbf {\bibinfo {volume}
  {20}},\ \bibinfo {pages} {L83} (\bibinfo {year} {1987})}\BibitemShut
  {NoStop}%
\bibitem [{\citenamefont {Tserkovnyak}\ and\ \citenamefont
  {Mecklenburg}(2008)}]{PhysRevB.77.134407}%
  \BibitemOpen
  \bibfield  {author} {\bibinfo {author} {\bibfnamefont {Y.}~\bibnamefont
  {Tserkovnyak}}\ and\ \bibinfo {author} {\bibfnamefont {M.}~\bibnamefont
  {Mecklenburg}},\ }\href {\doibase 10.1103/PhysRevB.77.134407} {\bibfield
  {journal} {\bibinfo  {journal} {Phys. Rev. B}\ }\textbf {\bibinfo {volume}
  {77}},\ \bibinfo {pages} {134407} (\bibinfo {year} {2008})}\BibitemShut
  {NoStop}%
\bibitem [{\citenamefont {Wong}\ and\ \citenamefont
  {Tserkovnyak}(2009)}]{PhysRevB.80.184411}%
  \BibitemOpen
  \bibfield  {author} {\bibinfo {author} {\bibfnamefont {C.~H.}\ \bibnamefont
  {Wong}}\ and\ \bibinfo {author} {\bibfnamefont {Y.}~\bibnamefont
  {Tserkovnyak}},\ }\href {\doibase 10.1103/PhysRevB.80.184411} {\bibfield
  {journal} {\bibinfo  {journal} {Phys. Rev. B}\ }\textbf {\bibinfo {volume}
  {80}},\ \bibinfo {pages} {184411} (\bibinfo {year} {2009})}\BibitemShut
  {NoStop}%
\bibitem [{\citenamefont {Onsager}(1931)}]{PhysRev.37.405}%
  \BibitemOpen
  \bibfield  {author} {\bibinfo {author} {\bibfnamefont {L.}~\bibnamefont
  {Onsager}},\ }\href {\doibase 10.1103/PhysRev.37.405} {\bibfield  {journal}
  {\bibinfo  {journal} {Phys. Rev.}\ }\textbf {\bibinfo {volume} {37}},\
  \bibinfo {pages} {405} (\bibinfo {year} {1931})}\BibitemShut {NoStop}%
\bibitem [{\citenamefont {Kim}\ \emph {et~al.}(2016)\citenamefont {Kim},
  \citenamefont {Takei},\ and\ \citenamefont
  {Tserkovnyak}}]{PhysRevB.93.020402}%
  \BibitemOpen
  \bibfield  {author} {\bibinfo {author} {\bibfnamefont {S.~K.}\ \bibnamefont
  {Kim}}, \bibinfo {author} {\bibfnamefont {S.}~\bibnamefont {Takei}}, \ and\
  \bibinfo {author} {\bibfnamefont {Y.}~\bibnamefont {Tserkovnyak}},\ }\href
  {\doibase 10.1103/PhysRevB.93.020402} {\bibfield  {journal} {\bibinfo
  {journal} {Phys. Rev. B}\ }\textbf {\bibinfo {volume} {93}},\ \bibinfo
  {pages} {020402} (\bibinfo {year} {2016})}\BibitemShut {NoStop}%
\bibitem [{\citenamefont {Abrikosov}(1957)}]{ABRIKOSOV1957199}%
  \BibitemOpen
  \bibfield  {author} {\bibinfo {author} {\bibfnamefont {A.~A.}\ \bibnamefont
  {Abrikosov}},\ }\href {\doibase https://doi.org/10.1016/0022-3697(57)90083-5}
  {\bibfield  {journal} {\bibinfo  {journal} {J. Phys. Chem. Solids}\ }\textbf
  {\bibinfo {volume} {2}},\ \bibinfo {pages} {199} (\bibinfo {year}
  {1957})}\BibitemShut {NoStop}%
\bibitem [{\citenamefont {Shockley}(1949)}]{6773080}%
  \BibitemOpen
  \bibfield  {author} {\bibinfo {author} {\bibfnamefont {W.}~\bibnamefont
  {Shockley}},\ }\href {\doibase 10.1002/j.1538-7305.1949.tb03645.x} {\bibfield
   {journal} {\bibinfo  {journal} {Bell Syst. Tech. J.}\ }\textbf {\bibinfo
  {volume} {28}},\ \bibinfo {pages} {435} (\bibinfo {year} {1949})}\BibitemShut
  {NoStop}%
\bibitem [{not()}]{notation}%
  \BibitemOpen
  \href@noop {} {}\bibinfo {note} {We use the normalization $\epsilon_{txy}=1,
  \epsilon_{xyz}=1$ and metric
  $g_{\mu\nu}=\text{diag}(1,-1,-1,-1)$.}\BibitemShut {Stop}%
\bibitem [{spi()}]{spinwave}%
  \BibitemOpen
  \href@noop {} {}\bibinfo {note} {Note that a finite vorticity requires
  nonlinear spin configurations and, in particular, cannot be carried by spin
  waves. To see this, consider a spin-wave texture
  $\vb{m}=\hat{\mathbf{z}}+\delta \vb{m}$, where $\delta m_x(\vb{r},t)\sim
  \cos(\vb{k}\cdot \vb{r}-\omega_k t)$ and $\delta m_y(\vb{r},t)\sim
  \sin(\vb{k}\cdot \vb{r}-\omega_kt)$. One can check that the corresponding
  vorticity has $\rho=0 \text{ and } \vb{j}=0$.}\BibitemShut {Stop}%
\bibitem [{\citenamefont {Nakahara}(2003)}]{homotopy}%
  \BibitemOpen
  \bibfield  {author} {\bibinfo {author} {\bibfnamefont {M.}~\bibnamefont
  {Nakahara}},\ }\href@noop {} {\emph {\bibinfo {title} {Geometry, Topology and
  Physics}}},\ \bibinfo {edition} {2nd}\ ed.\ (\bibinfo  {publisher} {CRC
  Press},\ \bibinfo {year} {2003})\BibitemShut {NoStop}%
\bibitem [{\citenamefont {Dubrovin}\ \emph {et~al.}(1985)\citenamefont
  {Dubrovin}, \citenamefont {Fomenko},\ and\ \citenamefont
  {Novikov}}]{geometry}%
  \BibitemOpen
  \bibfield  {author} {\bibinfo {author} {\bibfnamefont {B.}~\bibnamefont
  {Dubrovin}}, \bibinfo {author} {\bibfnamefont {A.}~\bibnamefont {Fomenko}}, \
  and\ \bibinfo {author} {\bibfnamefont {S.}~\bibnamefont {Novikov}},\
  }\href@noop {} {\emph {\bibinfo {title} {Modern Geometry--Methods and
  Applications: Part II: The Geometry and Topology of Manifolds}}}\ (\bibinfo
  {publisher} {Springer},\ \bibinfo {year} {1985})\BibitemShut {NoStop}%
\bibitem [{tor()}]{torque}%
  \BibitemOpen
  \href@noop {} {}\bibinfo {note} {We construct this torque phenomenologically,
  on symmetry grounds. Accordingly, the magnetic lead can be replaced with a
  nonmagnetic metal, in conjunction with an external magnetic field
  $\vb{B}=B\hat{\vb{z}}$. This setup would have the same structural and
  time-reversal symmetries. Note that this torque is generic, in the presence
  of an exchange coupling between electrons in the metal and magnetic moments
  of the insulator. In particular, it does not require microscopically the
  presence of a spin-orbit coupling.}\BibitemShut {Stop}%
\bibitem [{\citenamefont {Wigner}(1934)}]{Wigner:1934aa}%
  \BibitemOpen
  \bibfield  {author} {\bibinfo {author} {\bibfnamefont {E.}~\bibnamefont
  {Wigner}},\ }\href {\doibase 10.1103/PhysRev.46.1002} {\bibfield  {journal}
  {\bibinfo  {journal} {Phys. Rev.}\ }\textbf {\bibinfo {volume} {46}},\
  \bibinfo {pages} {1002} (\bibinfo {year} {1934})}\BibitemShut {NoStop}%
\bibitem [{\citenamefont {Kosterlitz}\ and\ \citenamefont
  {Thouless}(1973)}]{KT}%
  \BibitemOpen
  \bibfield  {author} {\bibinfo {author} {\bibfnamefont {J.~M.}\ \bibnamefont
  {Kosterlitz}}\ and\ \bibinfo {author} {\bibfnamefont {D.~J.}\ \bibnamefont
  {Thouless}},\ }\href {http://stacks.iop.org/0022-3719/6/i=7/a=010} {\bibfield
   {journal} {\bibinfo  {journal} {J. Phys. C}\ }\textbf {\bibinfo {volume}
  {6}},\ \bibinfo {pages} {1181} (\bibinfo {year} {1973})}\BibitemShut
  {NoStop}%
\bibitem [{\citenamefont {{Einstein}}(1905)}]{1905AnP...322..549E}%
  \BibitemOpen
  \bibfield  {author} {\bibinfo {author} {\bibfnamefont {A.}~\bibnamefont
  {{Einstein}}},\ }\href@noop {} {\bibfield  {journal} {\bibinfo  {journal}
  {Ann. Phys.}\ }\textbf {\bibinfo {volume} {322}},\ \bibinfo {pages} {549}
  (\bibinfo {year} {1905})}\BibitemShut {NoStop}%
\bibitem [{\citenamefont {H{\"a}nggi}\ \emph {et~al.}(1990)\citenamefont
  {H{\"a}nggi}, \citenamefont {Talkner},\ and\ \citenamefont
  {Borkovec}}]{RevModPhys.62.251}%
  \BibitemOpen
  \bibfield  {author} {\bibinfo {author} {\bibfnamefont {P.}~\bibnamefont
  {H{\"a}nggi}}, \bibinfo {author} {\bibfnamefont {P.}~\bibnamefont {Talkner}},
  \ and\ \bibinfo {author} {\bibfnamefont {M.}~\bibnamefont {Borkovec}},\
  }\href {\doibase 10.1103/RevModPhys.62.251} {\bibfield  {journal} {\bibinfo
  {journal} {Rev. Mod. Phys.}\ }\textbf {\bibinfo {volume} {62}},\ \bibinfo
  {pages} {251} (\bibinfo {year} {1990})}\BibitemShut {NoStop}%
\end{thebibliography}

%

\end{document}